# All-fiber broadband spectral acousto-optic modulation of a tubular-lattice hollow-core optical fiber

**Ricardo E. da Silva,[1,4*] Jonas H. Osório,[2,3] Gabriel L. Rodrigues,[2] David J. Webb,[4] Frédéric Gérôme,[5] Fetah Benabid,[5] Cristiano M. B. Cordeiro,[2] Marcos A.R. Franco[1]**

[1]*Institute for Advanced Studies (IEAv), São José dos Campos, 12228-001, Brazil*
[2]*Institute of Physics "Gleb Wataghin", University of Campinas (UNICAMP), Campinas, 13083-859, Brazil*
[3]*Department of Physics, Federal University of Lavras, Lavras, 37200-900, Brazil*
[4]*Aston Institute of Photonic Technologies (AIPT), Aston University, Birmingham, B4 7ET, UK*
[5]*GPPMM Group, XLIM Institute, UMR CNRS 7252, University of Limoges, Limoges, 87060, France*
**Corresponding author: r.da-silva@aston.ac.uk*

**We demonstrate a broadband acousto-optic notch filter based on a tubular-lattice hollow-core fiber for the first time. The guided optical modes are modulated by acoustically induced dynamic long-period gratings along the fiber. The device is fabricated employing a short interaction length (7.7 cm) and low drive voltages (10 V). Modulated spectral bands with 20 nm half-width and maximum depths greater than 60 % are achieved. The resonant notch wavelength is tuned from 743 to 1355 nm (612 nm span) by changing the frequency of the electrical signal. The results indicate a broader tuning range compared to previous studies using standard and hollow-core fibers. It further reveals unique properties for reconfigurable spectral filters and fiber lasers, pointing to the fast switching and highly efficient modulation of all-fiber photonic devices.**

All-fiber acousto-optic modulators (AOMs) have been successfully employed in reconfigurable spectral filters (rejection band, passband, and add-drop), heterodyne sensors, Q-switched and mode-locked fiber lasers [1–9]. In particular, flexural acoustic waves induce dynamic long-period gratings (LPGs) along an optical fiber, coupling power between the propagating optical modes. The resulting modulated notch depth and resonant wavelength are tuned by the acoustic amplitude and frequency, respectively. Thus, standing acoustic waves inducing amplitude-modulated spectral bands are suitable for mode-lock pulsed fiber lasers, controlling the laser gain threshold by adding losses in the laser cavity [7–9]. However, in standard optical fibers, the acoustic waves are mostly distributed over the fiber cladding reducing the overlap with the optical modes in the core.

AOMs using long fibers and high voltages have increased the acousto-optic interaction [1–3,5–7,9] while providing wide spectral tuning ranges employing distinct fiber designs: hollow-core photonic bandgap optical fibers (18 - 60 nm) [10,11], dispersion compensating fibers (36 – 80 nm) [1,2,6], few-mode fibers (100 nm) [5], single-mode standard fibers (500 nm) [3] and, solid-core photonic crystal fibers (1000 nm) [12]. The modulated full width at half maximum (FWHM) bands usually range from 1 to 7 nm [2,5,11,12].

Another opportunity explored in the literature is the utilization of fibers with reduced diameters. Nevertheless, a significant reduction of fiber diameter by means of etching or tapering might expose the modal properties to surface contamination, changing the original transmission spectrum. The fiber also becomes fragile degrading its stability. In addition, devices modulating too narrow spectral bands impose a strong filtering in the bandwidth of active media employed in fiber lasers, resulting in a small modulated spectral band of gain. Short fibers in AOMs increase the modulated bandwidth, which contributes to shortening the laser's pulse width [9]. Although broad modulated bands as wide as 18 nm can be achieved by applying simultaneously distinct electrical frequencies [1], the use of high voltages employing amplifiers to compensate for the decreasing acoustic amplitude and fiber lengths might increase the device's power consumption, heating, size, and cost.

Additionally, we have previously shown that suspended core fibers with large air holes increase the acoustically induced strain along the fiber, significantly strengthening the interaction of acoustic waves and Bragg gratings in the fiber core [4,8]. Here, we demonstrate that tubular-lattice hollow-core optical fibers (HCFs) further improve the modulation efficiency and spectral tuning without the propagation of acoustic waves inside the air-core. Thus, we numerically and experimentally investigate the modulation of the propagating optical modes by flexural acoustic waves in an HCF and report on the achieved performances.

Fig. 1(a) shows a detail of the single-ring tubular HCF we employ herein. The silica fiber is composed of a 30 μm diameter core, formed by 8 tubes of 10.7 μm in diameter and thickness of 300 nm. The fiber outer diameter is 200 μm. Fig. 1(b) shows the transmission spectrum of the HCF.

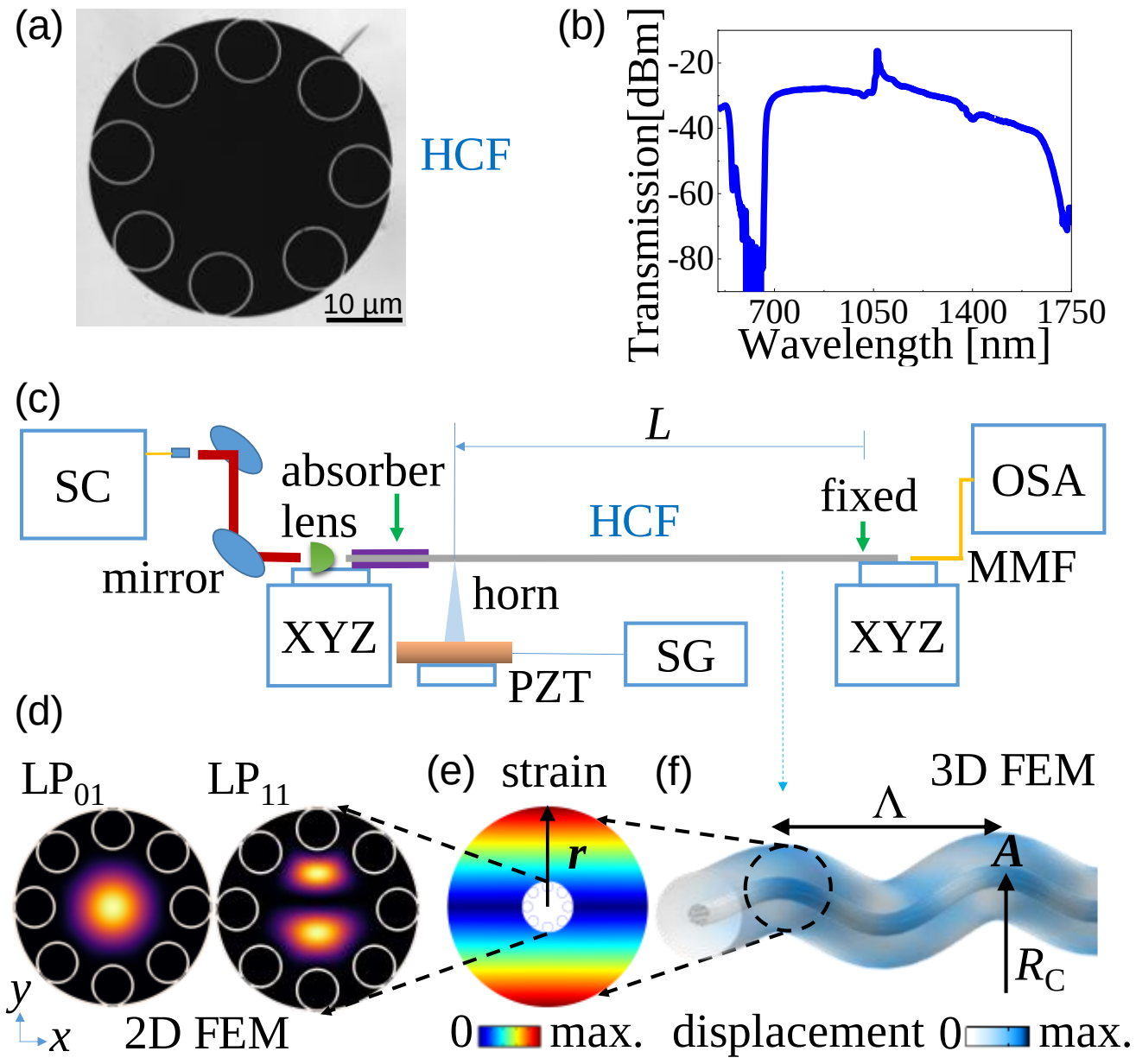

**Fig. 1.** Tubular lattice hollow-core fiber (HCF) with a (a) detail of the fiber core and (b) the transmission spectrum. (c) Illustration of the experimental setup and acousto-optic modulator. (d) 2D simulation of the power distribution of the fundamental optical mode $LP_{01}$ and higher-order optical mode $LP_{11}$ in the HCF. 3D simulation of the (e) strain distribution in the HCF cross-section caused by (f) the displacement a flexural acoustic wave of period $\Lambda$ and amplitude $A$. The induced curvature radius is $R_C$. The simulations are computed with the finite element method (FEM).

The HCF is fabricated according to the methodology described in [13]. The acousto-optic modulator is composed of an HCF segment, a piezoelectric transducer (PZT), and a capillary silica horn connected as illustrated in Fig. 1(c). The horn, PZT, and connections are similar to those employed in [14]. The output fiber end is fixed and the interaction length is $L$ = 7.7 cm. Acoustic waves outside the region $L$ are absorbed by the fiber coating. A sinusoidal electrical signal with a maximum voltage of 10 V is applied to the PZT in the frequency range from $f$ = 196 to 579 kHz using a signal generator (SG). The HCF's spectrum is characterized by using a supercontinuum source (SC NKT SK) and an optical spectrum analyzer (OSA Yokogawa AQ6370B). The coupling alignments are performed using XYZ micrometer stages, mirrors, and a coupling lens. The HCF output is butt-coupled directly to a multimode fiber (MMF) connected to the OSA.

The PZT generates standing flexural acoustic waves along the HCF, coupling the propagating fundamental mode $LP_{01}$ and the higher-order mode $LP_{11}$ at the resonant wavelength $\lambda_C$, satisfying the phase-matching condition as [1,5,10],

$$\Lambda = L_B = \frac{\lambda_C}{n_{01} - n_{11}}, \qquad (1)$$

where $\Lambda$ is the acoustic period, $L_B$ is the optical beat length, and $n_{01}$ and $n_{11}$ are the effective refractive indices of the modes $LP_{01}$ and $LP_{11}$ respectively. $\lambda_C$ is tuned by altering $\Lambda$ while changing the acoustic frequency $f$, as [1,5,10],

$$\Lambda = \left(\frac{\pi r c_{ext}}{f}\right)^{\frac{1}{2}}, \qquad (2)$$

where, $r$ is the fiber radius and, $c_{ext}$ = 5740 m/s is the extensional acoustic velocity [14]. The fiber bends polarized in the $y$ coordinate change the HCF's refractive index as [15],

$$n(x,y) = \left(n_0^2 + \frac{2y}{R_C}\right)^{\frac{1}{2}}, \qquad (3)$$

where, $n_0$ is the refractive index of the unbent fiber, and $R_C = 1/A\Omega^2$, is the curvature radius illustrated in Fig. 1(f). $\Omega = 2\pi/\Lambda$ is the acoustic wavenumber. The geometrical effect caused by the bends effectively increases the index, changing the optical path length of the guided modes in the air core.

We have simulated the HCF's modal properties and the frequency response of the flexural acoustic waves along $L$ = 7.7 cm. The HCF is modeled with the finite element method (FEM - COMSOL® software) employing the materials and methods described in [14]. Fig. 1(d) shows the simulated power distribution of the modes $LP_{01}$ and $LP_{11}$ (2D). Fig. 1(e), in turn, shows the modulus of the strain in the HCF cross-section caused by an arbitrary flexural acoustic wave of period $\Lambda$ and amplitude $A$ (Fig. 1(f) illustrates the fiber's displacements indicating the maxima (yale blue) and nodes (light gray) of the standing acoustic wave). Fig. 2(a) shows the simulated beat length $L_B$ of the modes $LP_{01}$ and $LP_{11}$ for the range of $\lambda$ = 750 - 1350 nm. The acoustic period $\Lambda$ is simulated from $f$ = 190 to 580 kHz and compared to $\Lambda$ calculated with Eq. (2) (Fig. 2(b)). Note that the agreement of $L_B$ and $\Lambda$ satisfies the phase-matching condition to tune resonance wavelengths $\lambda_C$ over a broad spectral range (> 600 nm). The phase-matching is not satisfied to other higher order modes because the larger difference of effective indices in Eq. (1) in the low loss transmission window of the HCF [13].

We measured two modulated notches in the transmission spectrum caused by acoustically induced birefringence in the HCF [2,3], which is further evaluated in the 2D simulation with Eq. (3). The curvature radius $R_C$ is estimated employing the acoustic period $\Lambda$ from the 3D simulation, considering the measured $\lambda_C$ and $f$. The acoustic amplitude is estimated from the experiment and average response of the PZT [14], changing from $A$ = 5 to 3.4 μm. The induced curvature radius is estimated from $R_C$ = 4.7 to 2.8 cm. Fig. 3 shows the effective index difference $\Delta n = n_{01} - n_{11}$ indicating the birefringence of the mode $LP_{11}$ compared to values for the HCF with no bends (unmodulated). The acoustically induced birefringence splits the effective index of $LP_{11}$ into the modes $LP_{11}^{Even}$ and $LP_{11}^{Odd}$.

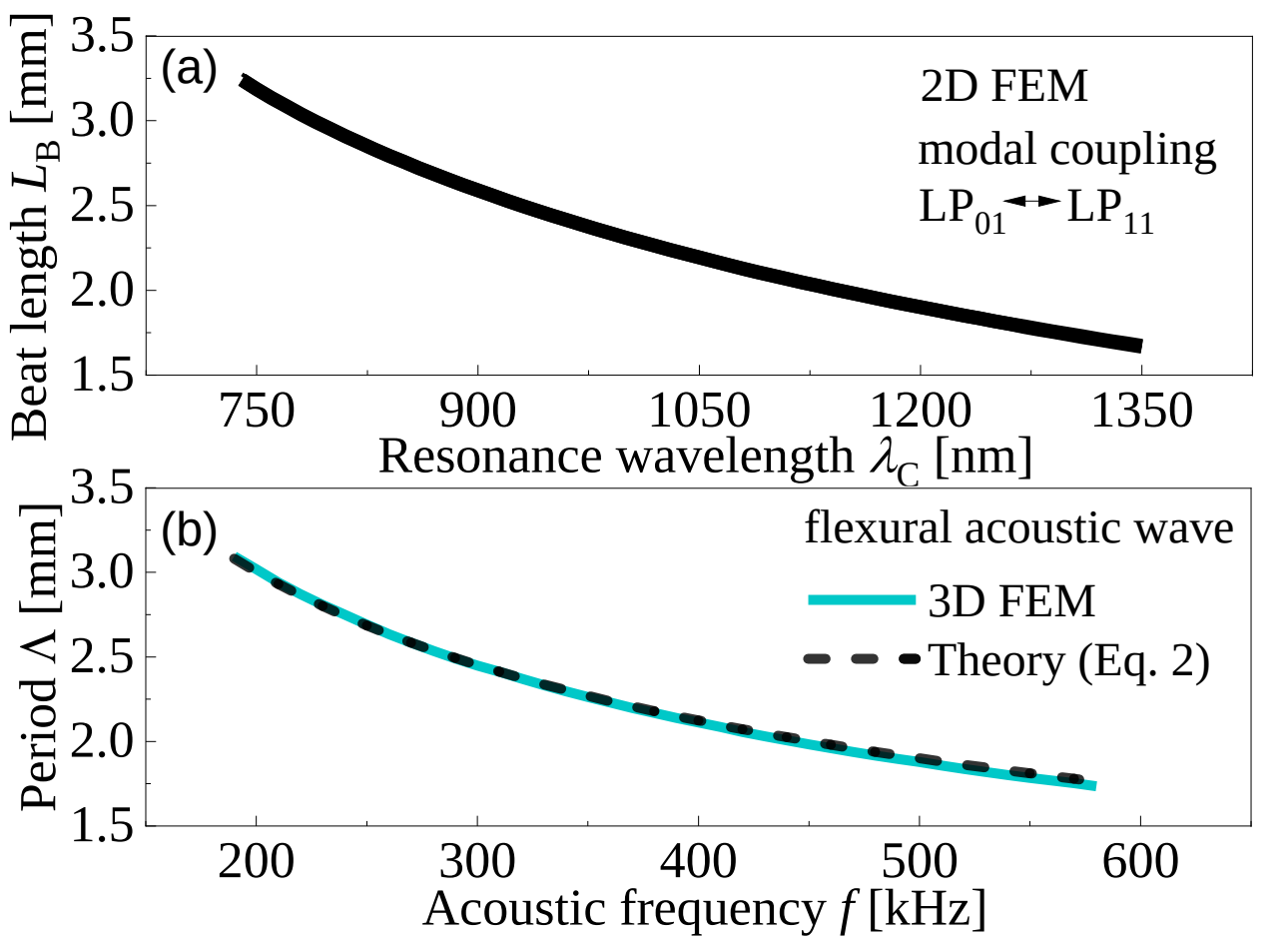

**Fig. 2.** (a) 2D FEM simulation of the optical beat length between the fundamental mode $LP_{01}$ and higher order mode $LP_{11}$ in the HCF. (b) 3D FEM simulation of the flexural acoustic waves indicating the period $\Lambda$ variation for the frequency range from $f$ = 190 to 580 kHz. The simulation is compared to the period $\Lambda$ calculated with Eq. (2).

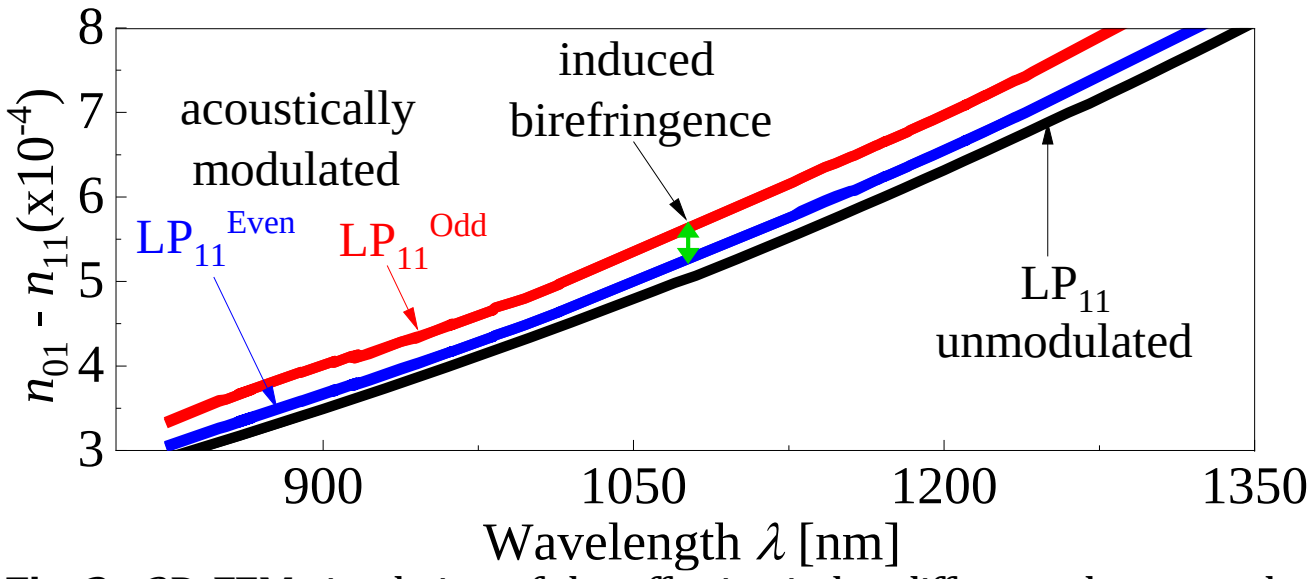

**Fig. 3.** 2D FEM simulation of the effective index difference between the fundamental and higher order modes $LP_{01}$ and $LP_{11}$ for the HCF with no bends (unmodulated) and acoustically modulated inducing birefringence between the even and odd $LP_{11}$ modes.

Consequently, two resonance bands are modulated in the HCF's transmission spectrum, as shown in Fig. 4(a). The average birefringence is $B_n$ = 3.7 x 10$^{-5}$, inducing an average resonances' separation of about $\lambda_{C2}$ - $\lambda_{C1}$ = 79 nm in the considered wavelength range.

The simulated power distribution of the even and odd $LP_{11}$ modes are shown as insets in Fig. 4(a). Fig. 4(b) shows the modulated spectrum for a voltage range from 2 to 10 V at $f$ = 400 kHz. The response from 1 to 10 V at $\lambda_{C2}$ = 1123 nm is shown as an inset in Fig. 4(b). Note that the slope (dB/V) changes slowly from 8 V tending to saturation by 10 V. It indicates higher modulation depths of the resonances because the OSA shows the average transmission of the unmodulated spectrum (acoustic wave off) and the spectrum modulated in amplitude at twice the acoustic frequency [9].

The average resonance bandwidth (FWHM) for the considered wavelength range is $\Delta\lambda$ = 20 nm, being significantly broader compared to previous studies employing similar interaction lengths or using multiple driven frequencies [2,5,11,12]. The modulated bandwidth depends on the interaction length $L$, but also on the fiber modal dispersion [3]. A short fiber length ($L$ = 7.7 cm) contributes to the broad modulated band but $\Delta\lambda$ can be further adjusted by appropriately choosing the HCF design to tailor the modal dispersion [13].

Samples of the modulated spectra of the mode $LP_{11}^{Odd}$ with increasing frequency are shown in Fig. 5(a). Although not shown, we mention that we achieved similar spectra with $LP_{11}^{Even}$ in a shorter wavelength range. The tuning of the resonance wavelength $\lambda_C$ is shown in Fig. 5(b). The reduced range of $\lambda_{C1}$ is caused by the decreasing displacement of the PZT with frequency [14], reducing the induced birefringence in the HCF. It also limited the tuning range to a maximum at $\lambda_C$ = 1355 nm. The limit at short wavelengths ($\lambda_C$ = 743 nm) is caused by the intrinsic high attenuation bands of the HCF [13,16] (Fig. 1(b)). Overall, both odd and even $LP_{11}$ resonances show an almost linear response ($r^2$ = 0.996), with tuning rates of 1.44 nm/kHz and 1.37 nm/kHz, respectively.

A detail of the resonance with the highest modulation depth of 3.9 dB (60% in linear scale) is seen in Fig. 5(a), which is comparable to results employing a standard optical fiber (SMF) under higher driven voltage of 18 V [9]. It indicates an overall 1.8x higher modulation efficiency for the HCF compared to the SMF, using a HCF outer diameter 1.6x larger. We conclude from the simulations and experiments that two important factors contribute to increasing the acousto-optic modulation efficiency in tubular lattice HCFs:

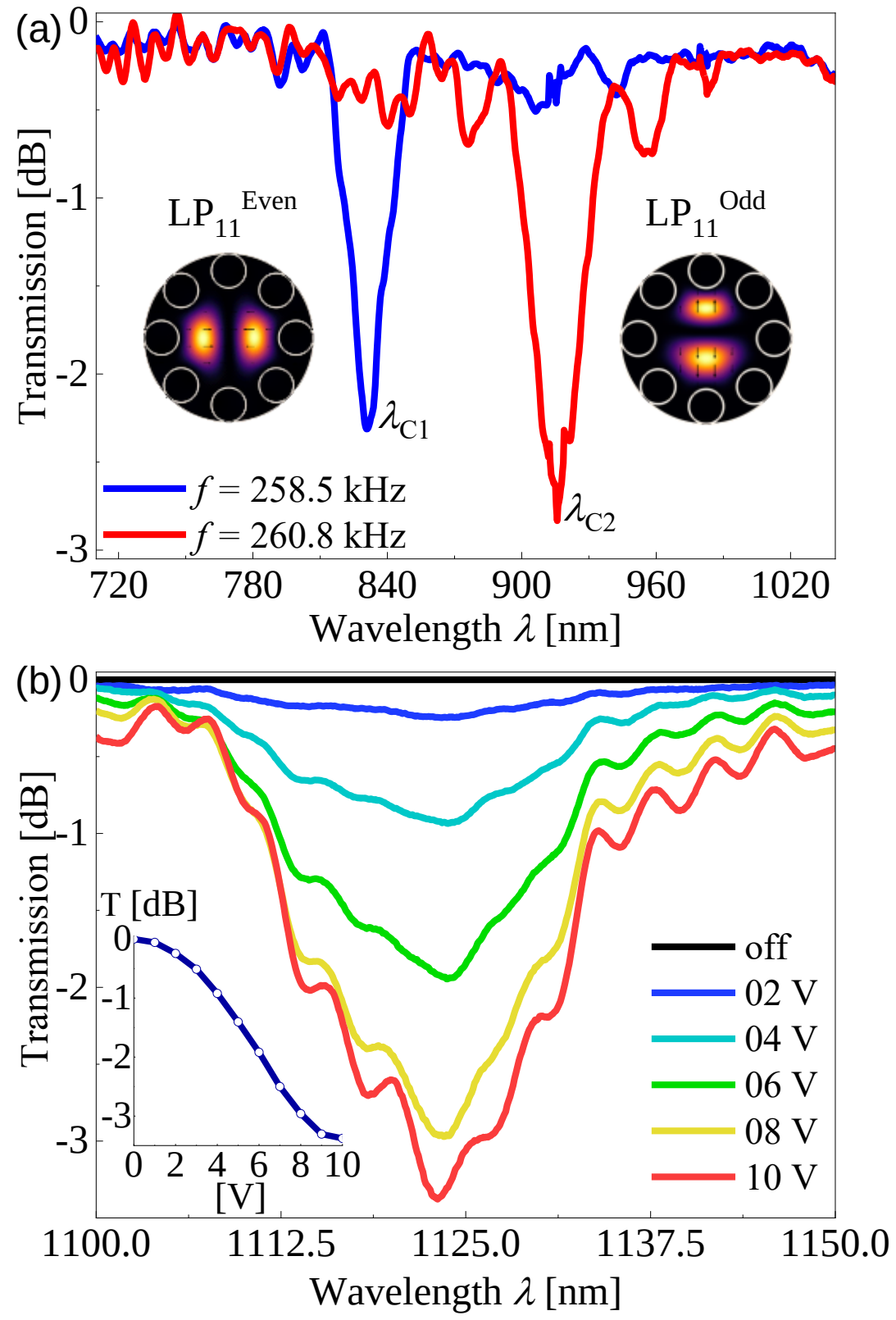

**Fig. 4.** (a) Acoustically modulated resonance notches in the transmission spectrum of the HCF. The resonant wavelengths $\lambda_{C1}$ and $\lambda_{C2}$ correspond respectively to the even and odd $LP_{11}$ modes, as illustrated by the 2D FEM simulation. (b) Unmodulated spectrum (acoustic wave off) and modulated spectra with increasing applied voltage up to 10 V. The detail shows the voltage response at $\lambda_{C2}$ = 1123 nm and $f$ = 400 kHz.

First, the reduction of the silica content in the fiber cross-section might increase the induced strain along the fiber, causing higher acoustic amplitudes compared to solid core fibers [4,8]. Second, the flexural acoustic waves are highly concentrated in the nanometric thickness tubes, which is exceptionally beneficial to modulating optical modes compared to solid core fibers. Note in Fig. 1(e) that the strain in HCF increases from zero ($r$ = 0) to a maximum at the fiber surface ($r$ = 100 μm). Considering $y = r$ in Eq. (3), the refractive index in the core increases accordingly with $r$. For an HCF and an SMF with the same diameter and curvature radius, the HCF would provide a higher change of the refractive index at the dielectric tubes compared to the SMF solid core, just because of the larger core radius ($r_{HCF}/r_{SMF}$ = 15 μm/4.1 μm = 3.6). It suggests the use of tubular lattice HCFs with large cores as a promising alternative to improve the modulation efficiency of acousto-optic devices. Higher modulation depths could also be achieved by the application of higher voltages to the PZT.

The switching response time of modulated resonances in AOMs is accurately approximated by the time the acoustic wave takes along the interaction length $L$, as $\tau = L/(2\sqrt{\pi r c_{ext} f})$ [1,3,6]. Considering the experimental parameters, $r$ = 100 μm, $c_{ext}$ = 5740 m/s, $L$ = 7.7 cm and $f$ = 579 kHz, the response time is estimated as 38 μs, which is relatively shorter compared to previous studies (57 μs) [1]. It indicates that the typical large diameter of HCFs is promising to decrease the switching response of AOMs while contributing to improve the mechanical stability.

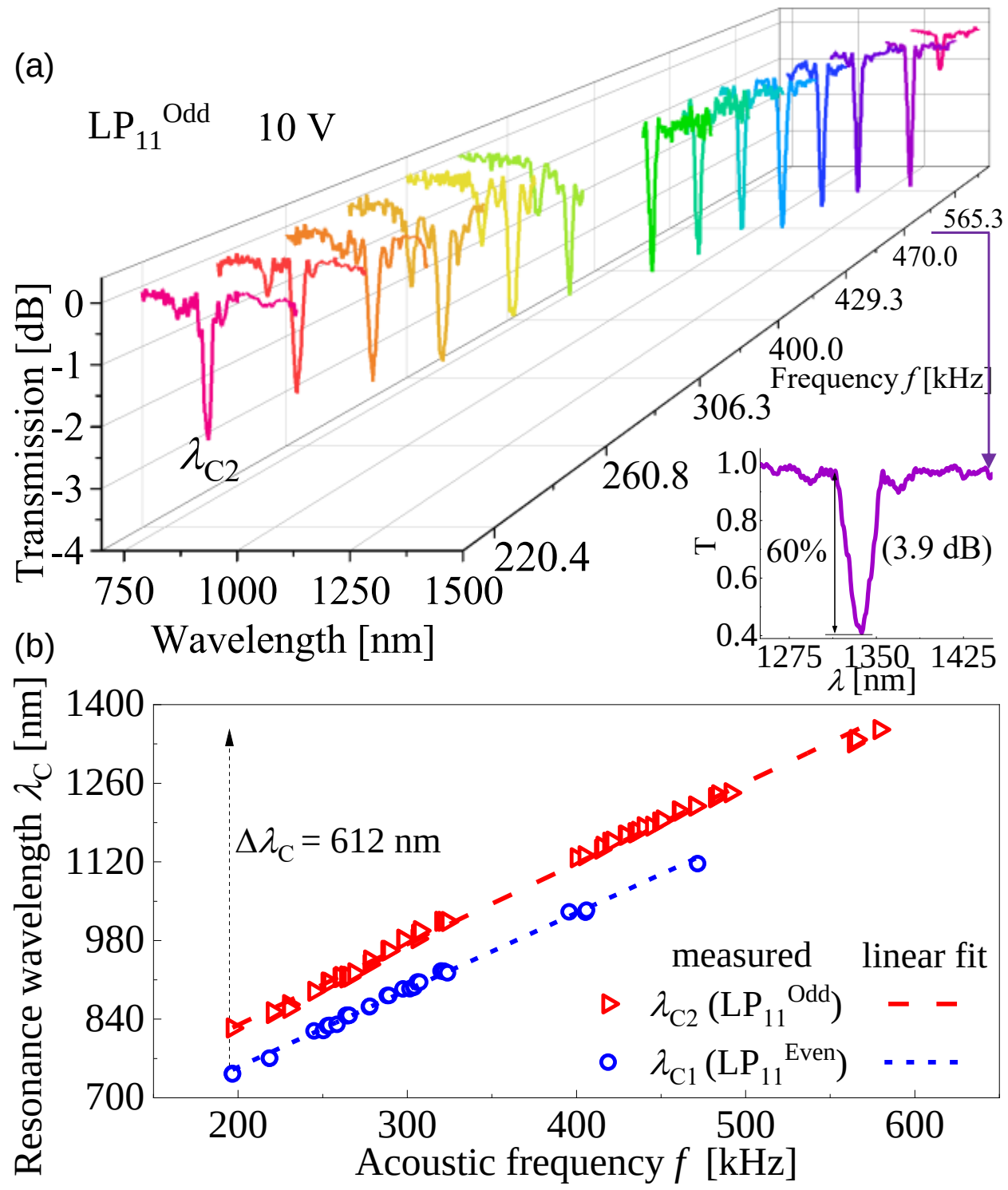

**Fig. 5.** (a) Modulated transmission spectra of the HCF showing the wavelength tuning of $\lambda_{C2}$ (mode $LP_{11}^{Odd}$) with increasing frequency. The detail shows the resonance with the highest modulation depth (60%). (b) Overall spectral tuning of the $LP_{11}$ resonances from $\lambda_C$ = 743 to 1355 nm in the frequency range from $f$ = 190 to 579 kHz.

Further advance of this study might suggest investigation of HCFs using other tubular dimensions and designs. In this way, modulated resonances with notch wavelengths and bandwidths being adjusted by the fiber geometry are expected. In addition, HCFs with large fiber air-cores are promising to increase the modulation depth. The large fiber diameter contributes to decreasing the response time of acousto-optic devices. Overall, tubular lattice HCFs additionally provide broader low-loss transmission windows compared to the photonic bandgap HCFs [11]. They enable great potential for acoustic modulation and spectral tuning over the wavelength range of important active media, such as erbium, ytterbium, neodymium, and YAG. The low nonlinearity and high damage threshold of HCFs allow in turn their application in high-power ultrafast fiber lasers [16] .

In summary, broadband spectral acousto-optic modulation of a tubular-lattice hollow-core fiber was experimentally demonstrated for the first time. The HCF's modal and acoustic properties were numerically investigated. The modulated spectral resonances achieved a maximum depth of 60% and an average bandwidth of 20 nm, which are useful for shortening the pulse width of mode-locked all-fiber lasers. The notch wavelengths were tuned in a broad range from 743 to 1355 nm (612 nm span). The HCF design and reduced interaction length contribute to design compact and fast acousto-optic devices, improving mechanical and optical stability, while enabling highly efficient acousto-optic modulation.

**Funding.** This project has received funding from grant #2022/10584-9, São Paulo Research Foundation (FAPESP), and grants 310650/2020-8 and 309989/2021-3, Conselho Nacional de Desenvolvimento Científico e Tecnológico (CNPq).

**Disclosures.** The authors declare no conflicts of interest.